\documentclass[letter,oldversion]{aa}
\usepackage{times}
\usepackage{mathptmx}
\usepackage{natbib}
\usepackage[]{fontenc}
\usepackage[latin1]{inputenc}
\usepackage[]{graphicx}

\begin{document}

\title{Dense gas without star formation: The kpc-sized turbulent
molecular disk in 3C326~N \thanks{Based on observations carried out
with the Very Large Telescope of ESO under program ID 385.B-0809.}}
\author{N.~P.~H.~Nesvadba\thanks{email:
nicole.nesvadba@ias.u-psud.fr}\inst{1}, F.~Boulanger\inst{1},
M.~D.~Lehnert\inst{2}, P.~Guillard\inst{3}, P.~Salome\inst{4}}
\institute{Institut d'Astrophysique Spatiale, CNRS, Universit\'e
Paris-Sud, 91405 Orsay, France
\and
GEPI, Observatoire de Paris, CNRS, Universit\'e Denis Diderot 5, Place
Jules Janssen, 92195 Meudon, France
\and
Spitzer Science Center, IPAC, California Institute of Technology, Pasadena, CA 92215, USA
\and
LERMA Observatoire de Paris, CNRS, 61, rue de l'Observatoire, 75014 Paris, France}

\authorrunning{Nesvadba et al.}
\titlerunning{The molecular disk in 3C326N}

\date{Received  / Accepted }

\abstract
{We report the discovery of a 3~kpc disk of few $10^9$ M$_{\odot}$ of
dense, warm H$_2$ in the nearby radio galaxy 3C326~N, which
shows no signs of on-going or recent star formation and 
falls a factor 60 below the Schmidt-Kennicutt law. VLT/SINFONI imaging
spectroscopy shows broad (FWHM$\sim$500 km s$^{-1}$) ro-vibrational
H$_2$ lines across all of the disk, with irregular profiles and line
ratios consistent with shocks. The ratio of turbulent  and
gravitational energy suggests that the gas is highly
turbulent and not gravitationally bound. In absence of the driving by
the jet, short turbulent dissipation times suggest the gas should
collapse rapidly and form stars, at odds with the recent
star-formation history. Motivated by hydrodynamic models of rapid
H$_2$ formation boosted by turbulent compression, we propose that the
molecules formed from diffuse atomic gas in the
turbulent jet cocoon. Since the gas is not self-gravitating, it cannot
form molecular clouds or stars while the jet is active, and is
likely to disperse and become atomic again after the nuclear activity
ceases. We speculate that very low star-formation rates are to be expected
under such conditions, provided that the large-scale turbulence sets
the gas dynamics in molecular clouds. Our results illustrate that
jets may create large molecular reservoirs as expected in 'positive
feedback' scenarios of AGN-triggered star formation, but that this
alone is not sufficient to trigger star formation.}


\maketitle

\section{Introduction} 
\label{sec:introduction}

The radio galaxy 3C326~N at z$=$0.1 is an excellent target to
investigate how the mechanical energy output of radio-loud AGN affects
the surrounding gas and star formation. In spite of a few $\times
10^9\ M_{\odot}$ of dense H$_2$, akin to LIRGs, the star-formation
rate \citep[SFR$\le$0.07 M$_{\odot}$ yr$^{-1}$][]{ogle07} is orders of
magnitude lower than expected from the Schmidt-Kennicutt law
\citep[][N10 hereafter]{nesvadba10}. With the low SFR, an X-ray faint
AGN \citep[$\log{{\cal L}_X}=40.6$ erg s$^{-1}$][]{ogle10}, and a
strong radio source (kinetic power $\log{{\cal L}_{kin}}\ge 44.6$ erg
s$^{-1}$, N10) 3C326~N provides a rare opportunity to clearly
disentangle the effects of the jet from those of star formation and
AGN radiation. It is one of the largest and oldest radio galaxies on
the sky, with 2 Mpc size, dynamical and spectral ages of
$6$-$20\times 10^7$ yrs \citep{willis78} and a radio core at 3~mm
(N10). 3C326~N is not an obvious member of a group or cluster, but has
a similarly old ($\ge 10$ Gyr), massive ($M_{stellar}=$few$\times
10^{11} M_{\odot}$) companion at a projected distance of $\sim 20$
kpc.

3C326~N is amongst the 30\% of nearby 3CR radio galaxies which have
bright line emission from warm molecular hydrogen observed with
Spitzer/IRS and line ratios consistent with shocks \citep{ogle10}. The
gas kinetic energy and line luminosities exceed the energy injection
rates from star formation and AGN radiation, making the deposition of
mechanical energy by the radio source the only plausible culprit
(N10). This energy injection does not only trigger an outflow, but
also heats the ambient gas through shocks, where the turbulent kinetic
energy roughly equals the bulk kinetic energy of the outflowing
gas. About half the molecular gas in 3C326~N is warm (T$>$100 K), 
a  fraction 100$\times$ greater than in star-forming galaxies, and
suggesting that mechanical heating by the jet could be an important
mode of suppressing the overall star formation in this galaxy.

We present deep imaging spectroscopy of the ro-vibrational H$_2$
emission lines at R$=$3000 and 0.7\arcsec\ spatial resolution, probing
the spatially resolved gas properties and kinematics, which
significantly enhances and complements the analysis of N10 which
relied on unresolved Spitzer/IRS spectroscopy at R=120 (2500 km
s$^{-1}$) with 5\arcsec\ spatial resolution. We find a kpc-sized
rotating, turbulent disk with broad complex line profiles, suggesting
that the mechanical energy deposition of the jet is not restricted to
a small part of the disk. We argue that finding such a disk is
difficult to reconcile with the recent star-formation history of
3C326~N unless the molecular gas formed in response to the turbulence
created by the radio jet, indicating a close symbiosis between jet and
molecular gas, but also suggesting that boosting the formation of
molecular gas in turbulent jet cocoons is not sufficient to trigger
star formation as often assumed in models of positive AGN
feedback. Throughout the paper we adopt a H$_0=$70 km s$^{-1}$,
$\Omega_M=$0.3, $\Omega_{\Lambda}=$0.7 cosmology.

\section{The molecular gas in 3C326~N}
\label{sec:observations}

\begin{figure}[t]
\centering
\includegraphics[width=0.5\textwidth]{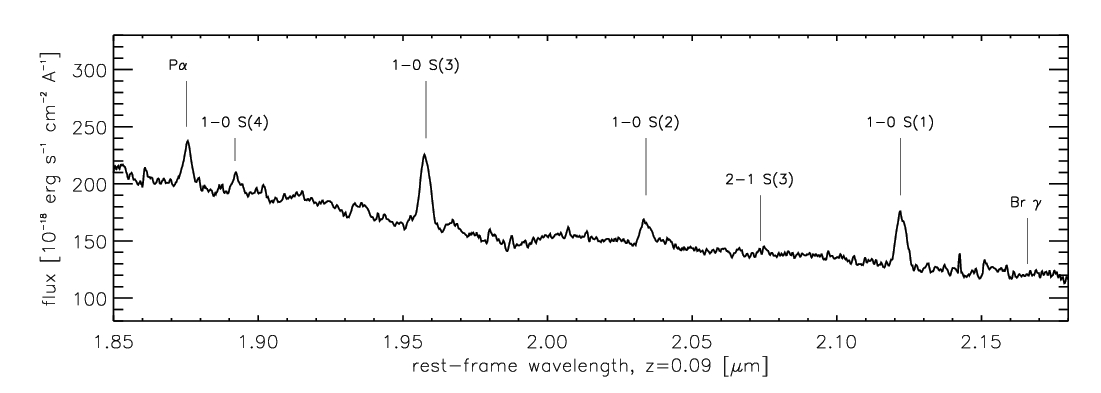}\\
\caption{Integrated K-band spectrum of 3C326~N. The properties of
labeled lines are listed in Table~\ref{tab:emlines}.}
\label{fig:intspec}
\end{figure}

\vspace{-1mm}
Data were obtained with the near-infrared imaging spectrograph SINFONI
\citep{bonnet04} at the Very Large Telescope of ESO with a total of
15,000 seconds of on-source observing time in the K-band at a
seeing-limited resolution of FWHM=0.75\arcsec$\times$0.65\arcsec.
Our data reduction and calibration methods are described by, e.g
\citet{nesvadba08a,nesvadba11a,nesvadba11b}.  The integrated spectrum
of 3C326~N (Fig.~\ref{fig:intspec}) is dominated by the ro-vibrational
H$_2$ lines H$_2$ 1-0 S(1) to S(4) (S(1)-S(4) hereafter). Pa$\alpha$
is very weak in comparison, Pa$\alpha$/S(3)$=$0.5
(Table~\ref{tab:emlines}).

Several arguments suggest this gas is heated by shocks, similar to the
pure-rotational H$_2$ and optical lines (N10). First, S(3)/Pa$\alpha$
and S(1)/Br$\gamma$ ratios are factors 10 greater than the typical
S(1)/Br$\gamma \sim 0.1-1.5$ in star-forming galaxies dominated by UV
heating \citep{puxley90}. Since we do not detect Br$\gamma$, we
estimate S(1)/Br$\gamma$=18-22 from the Pa$\alpha$ flux and a line
decrement Pa$\alpha$/Br$\gamma$=12-14, which exceeds ratios in
star-forming galaxies by a factor 10.  
Second, for star formation we expect
ratios of the H$_2$ 2-1/ H$_2$ 1-0 S(3) lines factors of a few greater
than the $3\sigma$ upper limit of 0.12 for S(1) and S(3) \citep[][
see \citealt{herrera11} for a similar analysis]{lepetit06}. A third
diagnostics is the [OI]$\lambda$6300/H$\alpha$ ratio as a function of
S(1)/Br$\gamma$ \citep{mouri89}. For the [OI]$\lambda$6300/H$\alpha$
ratio of N10 the S(1)/Br$\gamma$ ratio in 3C326~N exceeds that
expected from star formation by a factor 10. Forth, the H$_2$ line
luminosity of S(1)-S(3), $1\times 10^{41}$ erg s$^{-1}$, exceeds the
X-ray luminosity of the AGN and the mechanical luminosity of star
formation (N10), leaving shock heating by the jet as the only
plausible culprit.

\begin{figure}[th]
\centering
\includegraphics[width=0.5\textwidth]{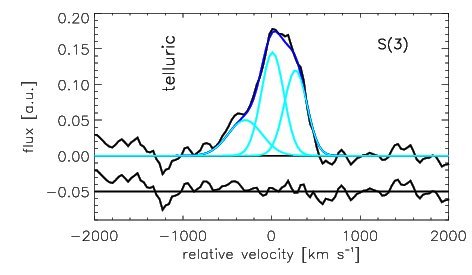}
\caption{Emission-line profile of H$_2$ 1-0 S(3) extracted from a
0.6\arcsec$\times$0.6\arcsec aperture on the Northern disk (black
line) fit with 3 Gaussian components (light blue line). The fit
residual is shown below.}
\label{fig:lineprofiles}
\end{figure}

Pa$\alpha$ and H$_2$ morphologies are disk-like and extend over
2.0\arcsec$\times$1.1\arcsec\ (3.2 kpc $\times$ 1.8 kpc) at
PA$_{morph}$=0$\pm$20$^{\circ}$.  All H$_2$ lines and Pa$\alpha$ are
broad, FWHM$\sim$400-700 km s$^{-1}$, down to the limit of our spatial
resolution. They have irregular profiles (Fig.~\ref{fig:lineprofiles})
that cannot be fit with single Gaussians. To map the kinematics, we
therefore measure the width, $W_p$, and central wavelength, $C_p$, of
each line at $p=$0.2, 0.5, 0.7, and 0.9$\times$ the line core
extracted from small apertures of 0.4\arcsec$\times0.4$\arcsec\
\citep{heckman81}. $W_{50}$ corresponds to the FWHM, and C$_{90}$
approximates the line core. For S(3) we also measure the asymmetry
parameter $A_{20}$, i.e., the shift of C$_{20}$ compared to C$_{90}$
\citep[Fig.~2 of][]{heckman81}. S(1) and Pa$\alpha$ have telluric
absorption features along their wings which makes measuring A$_{20}$
difficult, but their $C_{90}$ and $W_{50}$ give results consistent
with S(3).

\begin{figure}
\centering
\includegraphics[width=0.5\textwidth]{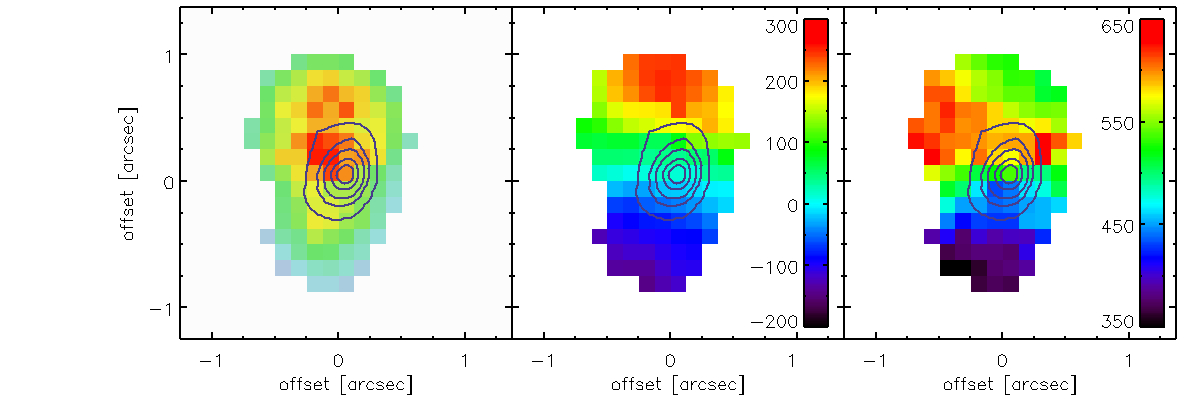}
\includegraphics[width=0.5\textwidth]{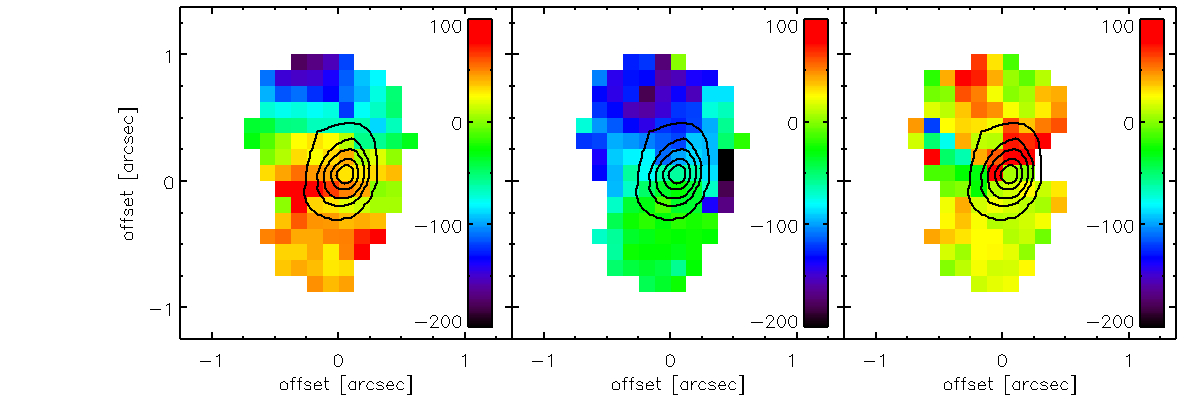}
\caption{{\it(top, left to right)} Maps of the line fluxes, relative
velocities, and FWHMs of H$_2$ 1-0 S(3)
in km s$^{-1}$. {\it(bottom, left to right)} Line asymmetry of
the 20-percentile for S(1), S(3), and Pa$\alpha$, respectively, in km
s$^{-1}$. Contours mark the continuum peak.}
\label{fig:maps}
\end{figure}

Velocities increase monotonically from SSW to NNE along
PA$=$345$^{\circ}\pm$15$^{\circ}$ (Fig.~\ref{fig:maps}) common to all
percentiles and all lines.  Lines are broadest NE of the nucleus
($W_{50,H2}\le$650 km s$^{-1}$, $W_{50,Pa}\le$750 km s$^{-1}$), and
more narrow in the south (300-400 km s$^{-1}$ for all lines).  A
Fourier analysis of non-circularities in the velocity map with
Kinemetry \citep{krajnovic06} suggests the velocity gradient is
indistinguishable from rotation with $PA_{kin}=340^{\circ}$,
inclination $i=35^{\circ}$ and a deprojected circular velocity of
$v_{c} = 1/2\ \Delta v_{proj}\ \csc i\ = 290$ km s$^{-1}$. This would
imply a dynamical mass $M_{dyn}=v_c^2\ R / G = 3\times 10^{10}\
M_{\odot}$ somewhat larger than the enclosed stellar mass of
$M_{stel}=2\times 10^{10}$ M$_{\odot}$. (The enclosed stellar mass is
derived from the total stellar mass of $3\times 10^{11}$ M$_{\odot}$
(N10) for a mass profile consistent with a Hubble law.) The
non-Gaussian line profiles suggest non-circular velocity components
which are probably smoothed out by the low spatial resolution. This 
could affect the velocity gradient.

The FWHMs are larger than the velocity gradient, hence the broad lines
cannot be artifacts of the velocity gradient and beam smearing, in
which case most of the broadening should be near the nucleus, contrary
to Fig.~\ref{fig:maps}.  The analysis of a toy data cube with constant
FWHM$=$400 km s$^{-1}$ and a gradient as observed suggests that beam
smearing accounts for $\le$15\% of the FWHM (similar to our 1$\sigma$
observational errors). We also used a toy data cube to quantify the
contribution of rotation to the CO(1-0) line measured by N10 which has
FWHM=350 km s$^{-1}$. We find a minimal intrinsic line width of 250 km
s$^{-1}$ assuming that the CO has the same velocity gradient as H$_2$
and Pa$\alpha$ and uniform surface brightness.

The small ratios of $v_c / \sigma=1.7-0.9$ (with velocity dispersion
$\sigma=FWHM/2.355=170-270$ km s$^{-1}$, Fig.~\ref{fig:maps}) imply an
ellipsoidal configuration rather than a thin disk. Broad lines
across all of the disk show that the mechanical energy of
the AGN affects the gas
globally. This differs strongly
from the scenario of jet-cloud interactions which are
confined to gas in small areas along the jet axis and is more akin 
to recent hydrodynamic models
of jet cocoons where radio-emitting plasma permeates inhomogeneous gas
disks along relatively low-density channels
\citep{sutherland07,wagner11}. 

The H$_2$ emission lines in 3C326~N have blue asymmetries which are
strongest in the north ($A_{20}=-150$ km s$^{-1}$, where turbulent
velocities are also largest). H$_2$ emission has velocities of up to
$-1000$ km s$^{-1}$ from $C_{90}$, well above the local escape
velocity, $v_{esc}\sim \sqrt{2}\ v_c \sim 400$ km s$^{-1}$ at the
largest H$_2$ radii.  $v_{esc}$ is also very similar to the lowest
FWHMs in the South. Fig.~\ref{fig:lineprofiles} suggests that of-order
10\% of the emission is from gas at $v>v_c$. Since the V(1-0) H$_2$
lines only probe the warmest, but not the bulk of the gas, this is
difficult to turn into a mass outflow rate $\dot{\rm M}$. If we assume
that the pure-rotational lines, which trace most of the mass, have a
similar profile \citep[as recently reported for AGN with
high-resolution Spitzer spectroscopy,][ Guillard et al. 2011 ApJ
submitted]{dasyra11}, we find $\sim 1\times 10^8$ M$_{\odot}$ of
entrained dense, warm H$_2$ in the wind.  For an outflow timescale of
$10^{7-8}$ yrs, this is consistent with or less than $\dot{\rm M}$
estimated from Na~D, $30-40$ M$_{\odot}$ yr$^{-1}$ (N10), which traces
neutral (atomic and/or molecular) gas down to lower column densities.

\vspace{-5mm}
\section{Jet-disk symbiosis} 
\label{sec:discussion}

\vspace{-1mm}
The gas in 3C326~N is in a kpc-sized, dense, thick disk. The broad
line widths and complex profiles are not consistent with pure rotation, but
require an additional source of turbulence which, following the energy
arguments of N10, can only be the radio source. A fraction of the line
emission is from gas at velocities above the local escape
velocity. The rotational time of this disk, $\tau_{rot}= 2 \pi R / v_c
= 3\times 10^7$ yrs, is shorter than the $6$-$20\times 10^7$ yrs age of
the radio source, and comparable to the depletion time of the gas
through the wind, $\tau_{wind}=M/\dot{M}\sim 5\times 10^7$ yrs. It is
also very similar to the dissipation time of turbulent energy 
$\tau_{diss}$=$R_{disk}/\sigma_{CO}=1.5\times 10^7$ yrs. The
similar time scales raise two major questions.

First, why do we find a fairly organized (albeit thick) disk, if the
depletion time through the wind is short? The mass outflow rate of N10
may be uncertain to about an order of magnitude, but even if it were a
factor 10 lower, much of the gas should have been removed in the
$2\times10^8$ yrs time span suggested by the oldest radio emission
(\S\ref{sec:introduction}). Finding a disk associated with an old
radio source like 3C326 requires that energy losses are
roughly balanced by the injection of energy, and the entrained by
cooling gas.  This would imply that a fraction of the outflowing gas
does not escape from the galaxy, but dissipates its kinetic energy,
stalls and rains back onto the disk in a cycle that may be akin to
'cold feedback' models of galaxy clusters \citep{ciotti91,
pizzolato05, brighenti06}.

Second, with $\tau_{diss}\sim \tau_{rot}$ the H$_2$ disk cannot be
stable in absence of the jet. The absence of a gas-rich merger
suggests the material accumulated gradually from mass return and
accretion over $\ga$few$\times 10^9$ yrs (N10), longer than the age of
the radio jet. Collisions between clouds would rapidly dissipate the
turbulent kinetic energy and the angular momentum of the disk in about
one to a few rotational times \citep[][]{verdoes06, pizzolato10},
consistent with our estimates. For the measured gas surface density
$\Sigma_{gas}=250\ M_{\odot}$ pc$^{-2}$, the Schmidt-Kennicutt law
implies SFR$=$4.5 M$_{\odot}$ yr$^{-1}$. This is $\sim$60$\times$
greater than the upper limit of SFR$=$0.07 M$_{\odot}$ yr$^{-1}$
\citep{ogle07} from Spitzer 70~$\mu$m imaging\footnote{for R=1.6
kpc. N10 found a smaller offset with a fiducial R=2.5 kpc}, and at
odds with the recent star-formation history of 3C326~N. To quantify
recent star formation we repeated the population synthesis analysis of
the inner 3\arcsec\ of 3C326~N from N10, this time adding small
fractions of light from young ($3$-$30 \times 10^7$ yrs) stellar
populations (YSPs) to the data. We find that atx SFR=4.5 M$_{\odot}$
yr$^{-1}$ stars must form for a few $\times 10^5$ yrs to produce
detectable signatures.  This is very short, hence, the absence of
recent star formation suggests that the gas cannot have lost turbulent
support for much longer than a free-fall time ($10^6$ yrs for N$=10^3$
cm$^{-3}$) in the last 300 Myr. Unless the jet has been nearly
continuously active during this time \citep[longer than the oldest
estimate of the jet age and unlike many other giant radio galaxies
with signs of repeated, shorter jet outbreaks,][]{schoenmakers00},
then why has this gas not been forming stars?

Resolving this paradox is possible if the molecules formed {\it after}
the onset of AGN activity, boosted by the turbulence in the jet
cocoon. Hydrodynamic simulations suggest that turbulent compression
can turn a diffuse inhomogeneous, largely atomic medium within few
$10^6$ yrs into a warm, largely molecular medium
\citep{glover07}. Molecules form in rather dense regions ($n\ge
1000$ cm$^{-3}$) and are transported to regions of lower density. The
total amount of H$_2$ on macro-scales is set by the dynamic
equilibrium of the rapid molecule formation and destruction on
micro-scales. This gas would not be gravitationally bound and be part
of a turbulent multiphase medium, where warm H$_2$ becomes a major gas
coolant \citep{guillard09}. 

In this scenario the molecular gas only survives the radio phase if it
forms self-gravitating clouds while pressurized by the
cocoon. Otherwise, with decreasing pressure H$_2$ formation rates
drop, and the dynamic equilibrium between molecule formation and
destruction is no longer maintained. The gas becomes again
atomic. Extended reservoirs of diffuse atomic gas are not unusual in
early-type galaxies \citep[][] {oosterloo07}. 
Dust lanes in radio
galaxies are often perpendicular to the jet \citep[][]{dekoff00}, and
have masses which depend on jet power \citep{deruiter02}, 
suggesting the ISM ``knows about the jet''.

\section{Turbulence in 3C326~N and star formation}
\label{sec:turbulence}
\vspace{-1mm}

To better understand how AGN and star formation in 3C326~N may be
related, we will now argue that turbulence governs the gas dynamics in
3C326~N down to sizes of molecular clouds. In this case finding low
star formation rates as observed may not be so surprising after
all. We quantify the ratio of turbulent to gravitational energy
through the virial parameter, $\alpha_{vir}=5\sigma_{cl}^2 (\pi\ G\
R_{cl}\ \Sigma_{gas})^{-1}$ \citep{bertoldi92}, where $\Sigma_{cl}$
and $R_{cl}$ are the mass surface density and radius of individual
clouds, respectively, and $G$ is the gravitational
constant. Gravitationally bound clouds have $\alpha_{vir}\le
1$. Hydrodynamic simulations suggest that column densities created by
turbulent motion cannot greatly exceed the mean column density
\citep[e.g.,][]{ostriker01}. We therefore use the measured
$\Sigma_{gas}=$250 $M_{\odot}$ pc$^{-2}$ (\S\ref{sec:observations}).

The H$_2$ line emission in 3C326~N is powered by the dissipation of
turbulent energy, which implies $L_{H2}/ M_{H2}= 3/2\ f_{H2}\
\sigma_{cl}^3/R_{cl}$ \citep{mckee07}, where $L_{H2}$ and $M_{H2}$ are
the $H_2$ luminosity and mass. N10 find $L_{H2}/M_{H2}\sim 0.06\
L_{\odot}/ M_{\odot}$. The correction factor $f_{H2}\la 1$ is
necessary because $H_2$ is the dominant, but not the only gas coolant
(see also Herrera et al. 2011, A\&A accepted). We adopt
$f_{H2}=0.5$. Turbulent velocities do not scale arbitrarily with size,
but approximately follow a power law $\sigma \propto R^{0.5}$, as
observed for molecular clouds in the Milky Way
\citep[e.g.,][]{larson81}, and seen in hydrodynamic simulations
including simulations of jet cocoons \citep[][]{krause07}.

\begin{figure}
\centering
\includegraphics[width=0.5\textwidth]{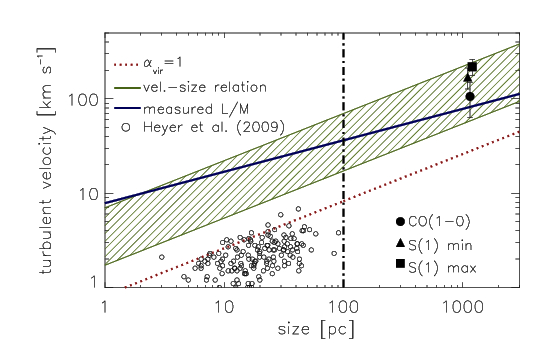}
\caption{Turbulent velocity as a function of size. Black symbols show
our data for CO, H$_2$1-0 lines, and Pa$\alpha$, respectively. The red
dotted line is for $\alpha_{vir}=1$ for a constant mass-surface density 
$\Sigma_{gas}=250$ M$_{\odot}$ pc$^{-2}$ and the green hatched region shows
the range implied by Larson's scaling law and the range of velocities
measured on kpc scales. Black circles show the molecular clouds of
\citet{heyer09}. These points fall near the $\alpha_{vir}=1$ relation
for gas surface densities $\sim 5$ smaller than that of 3C326~N.  The
horizontal dot-dashed line shows the maximal clump size of
$\sim$100~pc. See \S\ref{sec:turbulence} for details.}
\label{fig:alphavir}
\end{figure}

Fig.~\ref{fig:alphavir} illustrates that for all cloud sizes $\la 1$ kpc,
the observed L/M implies line widths at least 10$\times$ greater than
expected for $\alpha_{vir}=1$ (red line in Fig.~\ref{fig:alphavir}),
suggesting that the turbulent kinetic energy strongly exceeds the
gravitational energy on all relevant scales. This includes for
example the $\sim$100~pc scale at which disks akin to that of 3C326~N, but
without additional turbulence from the jet, would fragment
\citep[][dot-dashed vertical line in Fig.~\ref{fig:alphavir}]{escala08}, 
and also the cloud sizes of few 10s pc typically adopted 
in simulations of turbulent jet cocoons \citep{wagner11,antonuccio08}.

Fig.~\ref{fig:alphavir} shows also that turbulent velocities in
3C326~N are likely much greater than those in typical molecular clouds
in the Milky Way \citep[][grey circles in 
Fig.~\ref{fig:alphavir}]{heyer09}. This may hold the key to the low
star-formation rates. Local density enhancements owing to turbulent
compression may enable star formation in clouds with $\alpha_{vir}\ga
1$, but the star formation efficiency decreases rapidly with
increasing $\alpha_{vir}$ \citep[e.g.,][]{krumholz05, padoan11}.

These arguments rely on the assumption that Larson's scaling
relationship approximately holds for 3C326~N. Resolving scales of
$\sim$100~pc will only become possible once ALMA reaches its full
capabilities, however, we already see in Fig.~\ref{fig:alphavir} that
Larson's scaling law (green hatched area) normalized to the range of
velocities at kpc scales falls within the range of turbulence required
to account for the observed H$_2$ luminosity.\vspace{-3mm}
\section{Concluding remarks}
Our results suggest that AGN can create large reservoirs of dense,
rapidly cooling molecular gas, as is suggested by models of
AGN-triggered star formation \citep[e.g.,][]{mellema02, fragile04,
silk09}, but this does not {\it per se} seem sufficient to form stars.
The turbulent energy injected by the jet seems capable to keep
significant fractions of the gas warm (T$>$150 K) and gravitationally
unbound, making star formation very inefficient. 

3C326~N is certainly an extreme example of a gas-rich radio galaxy
with a very low star-formation efficiency, however, it is not
unique. 30\% of nearby 3CR radio galaxies have bright line emission
from warm H$_2$ \citep{ogle10}, and emission-line diagnostics suggest
that much of this gas could be heated mechanically through shocks (N10). 
This includes galaxies like 3C293, which have on-going star formation,
albeit not at the level expected from their H$_2$ reservoirs
\citep{papadopoulos10}. It will be interesting to investigate in more
depth how star formation and molecular gas kinematics are related in
these cases.

\begin{acknowledgements}
We are very grateful to the staff at Paranal Observatory for carrying
out the observations. Without the continuous, excellent work of the ESO 
staff and fellows, our analysis would not have been possible. We thank the 
referee for comments that helped improve the manuscript. 

\vspace{-4mm}
\end{acknowledgements}

\bibliographystyle{aa}
\bibliography{H2RG}

\begin{thebibliography}{42}
\expandafter\ifx\csname natexlab\endcsname\relax\def\natexlab#1{#1}\fi

\bibitem[{{Antonuccio-Delogu} \& {Silk}(2008)}]{antonuccio08}
{Antonuccio-Delogu}, V. \& {Silk}, J. 2008, \mnras, 389, 1750

\bibitem[{{Bertoldi} \& {McKee}(1992)}]{bertoldi92}
{Bertoldi}, F. \& {McKee}, C.~F. 1992, \apj, 395, 140

\bibitem[{{Bonnet} {et~al.}(2004){Bonnet}, {Abuter}, {Baker}, {Bornemann},
  {Brown}, {Castillo}, {Conzelmann}, {Damster}, {Davies}, {Delabre},
  {Donaldson}, {Dumas}, {Eisenhauer}, {Elswijk}, {Fedrigo}, {Finger},
  {Gemperlein}, {Genzel}, {Gilbert}, {Gillet}, {Goldbrunner}, {Horrobin}, {Ter
  Horst}, {Huber}, {Hubin}, {Iserlohe}, {Kaufer}, {Kissler-Patig}, {Kragt},
  {Kroes}, {Lehnert}, {Lieb}, {Liske}, {Lizon}, {Lutz}, {Modigliani}, {Monnet},
  {Nesvadba}, {Patig}, {Pragt}, {Reunanen}, {R{\"o}hrle}, {Rossi}, {Schmutzer},
  {Schoenmaker}, {Schreiber}, {Stroebele}, {Szeifert}, {Tacconi}, {Tecza},
  {Thatte}, {Tordo}, {van der Werf}, \& {Weisz}}]{bonnet04}
{Bonnet}, H., {Abuter}, R., {Baker}, A., {et~al.} 2004, The Messenger, 117, 17

\bibitem[{{Brighenti} \& {Mathews}(2006)}]{brighenti06}
{Brighenti}, F. \& {Mathews}, W.~G. 2006, \apj, 643, 120

\bibitem[{{Ciotti} {et~al.}(1991){Ciotti}, {D'Ercole}, {Pellegrini}, \&
  {Renzini}}]{ciotti91}
{Ciotti}, L., {D'Ercole}, A., {Pellegrini}, S., \& {Renzini}, A. 1991, \apj,
  376, 380

\bibitem[{{Dasyra} \& {Combes}(2011)}]{dasyra11}
{Dasyra}, K.~M. \& {Combes}, F. 2011, ArXiv e-prints

\bibitem[{{de Koff} {et~al.}(2000){de Koff}, {Best}, {Baum}, {Sparks},
  {R{\"o}ttgering}, {Miley}, {Golombek}, {Macchetto}, \& {Martel}}]{dekoff00}
{de Koff}, S., {Best}, P., {Baum}, S.~A., {et~al.} 2000, \apjs, 129, 33

\bibitem[{{de Ruiter} {et~al.}(2002)}]{deruiter02}
{de Ruiter}, H.~R. {et~al.} 2002, \aap, 396, 857

\bibitem[{{Escala} \& {Larson}(2008)}]{escala08}
{Escala}, A. \& {Larson}, R.~B. 2008, \apjl, 685, L31

\bibitem[{{Fragile} {et~al.}(2004){Fragile}, {Murray}, {Anninos}, \& {van
  Breugel}}]{fragile04}
{Fragile}, P.~C., {Murray}, S.~D., {Anninos}, P., \& {van Breugel}, W. 2004,
  \apj, 604, 74

\bibitem[{{Glover} \& {Mac Low}(2007)}]{glover07}
{Glover}, S.~C.~O. \& {Mac Low}, M. 2007, \apj, 659, 1317

\bibitem[{{Guillard} {et~al.}(2009){Guillard}, {Boulanger}, {Pineau Des
  For{\^e}ts}, \& {Appleton}}]{guillard09}
{Guillard}, P., {Boulanger}, F., {Pineau Des For{\^e}ts}, G., \& {Appleton},
  P.~N. 2009, \aap, 502, 515

\bibitem[{{Heckman} {et~al.}(1981)}]{heckman81}
{Heckman}, T.~M. {et~al.} 1981, \apj, 247, 403

\bibitem[{{Herrera} {et~al.}(2011){Herrera}, {Boulanger}, \&
  {Nesvadba}}]{herrera11}
{Herrera}, C.~N., {Boulanger}, F., \& {Nesvadba}, N.~P.~H. 2011, ArXiv e-prints

\bibitem[{{Heyer} {et~al.}(2009){Heyer}, {Krawczyk}, {Duval}, \&
  {Jackson}}]{heyer09}
{Heyer}, M., {Krawczyk}, C., {Duval}, J., \& {Jackson}, J.~M. 2009, \apj, 699,
  1092

\bibitem[{{Krajnovi{\'c}} {et~al.}(2006)}]{krajnovic06}
{Krajnovi{\'c}}, D. {et~al.} 2006, \mnras, 366, 787

\bibitem[{{Krause} \& {Alexander}(2007)}]{krause07}
{Krause}, M. \& {Alexander}, P. 2007, \mnras, 376, 465

\bibitem[{{Krumholz} \& {McKee}(2005)}]{krumholz05}
{Krumholz}, M.~R. \& {McKee}, C.~F. 2005, \apj, 630, 250

\bibitem[{{Larson}(1981)}]{larson81}
{Larson}, R.~B. 1981, \mnras, 194, 809

\bibitem[{{Le Petit} {et~al.}(2006){Le Petit}, {Nehm{\'e}}, {Le Bourlot}, \&
  {Roueff}}]{lepetit06}
{Le Petit}, F., {Nehm{\'e}}, C., {Le Bourlot}, J., \& {Roueff}, E. 2006, \apjs,
  164, 506

\bibitem[{{McKee} \& {Ostriker}(2007)}]{mckee07}
{McKee}, C.~F. \& {Ostriker}, E.~C. 2007, \araa, 45, 565

\bibitem[{{Mellema} {et~al.}(2002){Mellema}, {Kurk}, \&
  {R{\"o}ttgering}}]{mellema02}
{Mellema}, G., {Kurk}, J.~D., \& {R{\"o}ttgering}, H.~J.~A. 2002, \aap, 395,
  L13

\bibitem[{{Mouri} {et~al.}(1989){Mouri}, {Taniguchi}, {Kawara}, \&
  {Nishida}}]{mouri89}
{Mouri}, H., {Taniguchi}, Y., {Kawara}, K., \& {Nishida}, M. 1989, \apjl, 346,
  L73

\bibitem[{{Nesvadba} {et~al.}(2011{\natexlab{a}}){Nesvadba}, {De Breuck},
  {Lehnert}, \& {Best}}]{nesvadba11a}
{Nesvadba}, N., {De Breuck}, C., {Lehnert}, M., \& {Best}. 2011{\natexlab{a}},
  \aap, 525, A43+

\bibitem[{{Nesvadba} {et~al.}(2010){Nesvadba}, {Boulanger}, {Salom{\'e}},
  {Guillard}, {Lehnert}, {Ogle}, {Appleton}, {Falgarone}, \& {Pineau Des
  Forets}}]{nesvadba10}
{Nesvadba}, N.~P.~H., {Boulanger}, F., {Salom{\'e}}, P., {et~al.} 2010, \aap,
  521, A65+

\bibitem[{{Nesvadba} {et~al.}(2008){Nesvadba}, {Lehnert}, {Davies}, {Verma}, \&
  {Eisenhauer}}]{nesvadba08a}
{Nesvadba}, N.~P.~H., {Lehnert}, M.~D., {Davies}, R.~I., {Verma}, A., \&
  {Eisenhauer}, F. 2008, \aap, 479, 67

\bibitem[{{Nesvadba} {et~al.}(2011{\natexlab{b}}){Nesvadba}, {Polletta},
  {Lehnert}, {Bergeron}, {De Breuck}, {Lagache}, \& {Omont}}]{nesvadba11b}
{Nesvadba}, N.~P.~H., {Polletta}, M., {Lehnert}, M.~D., {et~al.}
  2011{\natexlab{b}}, ArXiv e-prints

\bibitem[{{Ogle} {et~al.}(2007){Ogle}, {Antonucci}, {Appleton}, \&
  {Whysong}}]{ogle07}
{Ogle}, P., {Antonucci}, R., {Appleton}, P.~N., \& {Whysong}, D. 2007, \apj,
  668, 699

\bibitem[{{Ogle} {et~al.}(2010){Ogle}, {Boulanger}, {Guillard}, {Evans},
  {Antonucci}, {Appleton}, {Nesvadba}, \& {Leipski}}]{ogle10}
{Ogle}, P., {Boulanger}, F., {Guillard}, P., {et~al.} 2010, \apj, 724, 1193

\bibitem[{{Oosterloo} {et~al.}(2007){Oosterloo}, {Morganti}, {Sadler}, {van der
  Hulst}, \& {Serra}}]{oosterloo07}
{Oosterloo}, T.~A., {Morganti}, R., {Sadler}, E.~M., {van der Hulst}, T., \&
  {Serra}, P. 2007, \aap, 465, 787

\bibitem[{{Ostriker} {et~al.}(2001){Ostriker}, {Stone}, \&
  {Gammie}}]{ostriker01}
{Ostriker}, E.~C., {Stone}, J.~M., \& {Gammie}, C.~F. 2001, \apj, 546, 980

\bibitem[{{Padoan} \& {Nordlund}(2011)}]{padoan11}
{Padoan}, P. \& {Nordlund}, {\AA}. 2011, \apj, 730, 40

\bibitem[{{Papadopoulos} {et~al.}(2010){Papadopoulos}, {van der Werf}, {Isaak},
  \& {Xilouris}}]{papadopoulos10}
{Papadopoulos}, P.~P., {van der Werf}, P., {Isaak}, K., \& {Xilouris}, E.~M.
  2010, \apj, 715, 775

\bibitem[{{Pizzolato} \& {Soker}(2005)}]{pizzolato05}
{Pizzolato}, F. \& {Soker}, N. 2005, \apj, 632, 821

\bibitem[{{Pizzolato} \& {Soker}(2010)}]{pizzolato10}
{Pizzolato}, F. \& {Soker}, N. 2010, \mnras, 408, 961

\bibitem[{{Puxley} {et~al.}(1990){Puxley}, {Hawarden}, \&
  {Mountain}}]{puxley90}
{Puxley}, P.~J., {Hawarden}, T.~G., \& {Mountain}, C.~M. 1990, \apj, 364, 77

\bibitem[{{Schoenmakers} {et~al.}(2000){Schoenmakers}, {de Bruyn},
  {R{\"o}ttgering}, {van der Laan}, \& {Kaiser}}]{schoenmakers00}
{Schoenmakers}, A.~P., {de Bruyn}, A.~G., {R{\"o}ttgering}, H.~J.~A., {van der
  Laan}, H., \& {Kaiser}, C.~R. 2000, \mnras, 315, 371

\bibitem[{{Silk} \& {Norman}(2009)}]{silk09}
{Silk}, J. \& {Norman}, C. 2009, \apj, 700, 262

\bibitem[{{Sutherland} \& {Bicknell}(2007)}]{sutherland07}
{Sutherland}, R.~S. \& {Bicknell}, G.~V. 2007, \apjs, 173, 37

\bibitem[{{Verdoes Kleijn} {et~al.}(2006){Verdoes Kleijn}, {van der Marel}, \&
  {Noel-Storr}}]{verdoes06}
{Verdoes Kleijn}, G.~A., {van der Marel}, R.~P., \& {Noel-Storr}, J. 2006, \aj,
  131, 1961

\bibitem[{{Wagner} \& {Bicknell}(2011)}]{wagner11}
{Wagner}, A.~Y. \& {Bicknell}, G.~V. 2011, \apj, 728, 29

\bibitem[{{Willis} \& {Strom}(1978)}]{willis78}
{Willis}, A.~G. \& {Strom}, R.~G. 1978, \aap, 62, 375

\end{thebibliography}

\begin{table}
\begin{center} 
\tiny{
\begin{tabular}{lcccc} 
\hline 
 Line & $\lambda_0$ & $\lambda_{obs}$ & FWHM & flux \\
\hline
     & [$\mu$m]     & [$\mu$m] & [km s$^{-1}$] & [$10^{-15}$ erg s$^{-1}$ cm$^{-2}$]\\
\hline
\hline
Pa$\alpha$     & 1.875 & 2.0443$\pm$0.0004 & 516$\pm$44 & 1.2$\pm$0.14\\
H$_2$ 1-0 S(1) & 2.122 & 2.3132$\pm$0.0005 & 534$\pm$32 & 1.8$\pm$0.13\\
H$_2$ 1-0 S(2) & 2.034 & 2.2170$\pm$0.0003 & 563$\pm$45 & 0.7$\pm$0.05 \\
H$_2$ 1-0 S(3) & 1.958 & 2.1339$\pm$0.0004 & 594$\pm$41 & 2.2$\pm$0.2 \\
H$_2$ 1-0 S(4) & 1.892 & 2.0624$\pm$0.0004 & 475$\pm$45 & 0.8$\pm$0.05 \\
\hline
Br$\gamma$     & 2.166 &                   &            & $<$0.26 (3$\sigma$)\\
H$_2$ 2-1 S(3) & 2.074 &                   &            & $<$0.24 (3$\sigma$) \\ 
\hline
\end{tabular}
\caption{Line properties derived from the ``stacked'' spectrum, i.e.,
the integrated spectrum with the large-scale velocity gradient
removed. We also list the 3$\sigma$ upper limit for Br$\gamma$, and
H$_2$ 2-1 S(3), which are not detected. }
\label{tab:emlines}}
\end{center}
\end{table}

\end{document}